\documentclass[a4paper,12pt]{article}

\usepackage{graphicx}
\usepackage{epsfig}

\usepackage{amssymb}
\begin{document}

\title{Random matrix approach in search for weak signals immersed in background  noise}

\author{D. Grech, J. Mi\'{s}kiewicz \\
Institute of Theoretical Physics\\
University of Wroc{\l}aw, PL-50-204 Wroc{\l}aw, Poland}

\date{}
\maketitle

\begin{abstract}

We present new, original and alternative method for searching signals coded in noisy data. The method is based on the properties of random matrix eigenvalue spectra. First, we describe general ideas and support them with results of numerical simulations for basic periodic signals immersed in artificial stochastic noise. Then, the main effort is put to examine the strength of a new method in investigation of data content taken from the real astrophysical  NAUTILUS detector, searching for the presence of gravitational waves. Our method discovers some previously unknown problems with data aggregation in this experiment.
We provide also the results of new method applied to the entire respond signal from ground based detectors in future experimental activities with reduced background noise level.
We indicate good performance of our method what makes it a positive predictor for further applications in many areas.

\end{abstract}

\section{Introduction}
One often encounters a problem if in the given noisy experimental data some repeatable weak signal is hidden. If such signal is really very weak, even its existence in stochastic background data is difficult to confirm, not mentioning the challenge to describe its detailed properties. This problem is relevant not only in detector data analysis in many fields of physics but it is also important in other areas of human activity where data transmission, data collection and data processing is involved -- just to mention: telecommunication, electronics, computer science, genetics, acoustics, astronomy, etc.

The search for long-memory effects in the measured  background noise
is essential in proper determination of the conceivable
periodic or just repeatable signal immersed in this noise. Such search is also the
first step to reveal and then to separate other sources of
long-memory data which, if added to the measured background noise, may finally change its nature.

In this report we introduce the novel method to
examine the presence of long-memory and autocorrelations
in time series. The method is based on properties of random matrices (RM)  eigenvalue spectra [1-3], however, we use RM approach in the new context.

Usually, one applies RM to multidimensional time series of data, seeking for correlations between  various $1$-dimensional subseries of these time series [4-6]. This technique analyzes eigenvalues of the correlation matrix between $1$-dimensional subseries with the corresponding eigenvalues of Wishart matrix [7,8]. In our approach, we propose  to analyze eigenvalue spectra for matrix entries built entirely from increments of $1$-dimensional time series with no correspondence to Wishart correlation matrix. The construction runs as follows.

Let $\{x_i\}, (i=1,...,N+1)$ is the $1$-dimensional cumulated time series with very large $N$. The time series of its increments is thus $\{\Delta x_i\}$ with $\Delta x_i = x_{i+1}-x_{i}$. We divide $\{\Delta x_i\}$ into $[N/L]$ non-overlapping subseries $s_k = \{\Delta x_{(k-1)L+1}, ..., \Delta x_{kL}\}$ of length $L$ each $(k=1, ..., [N/L])$. Every subseries is then renormalized according to:
\begin{equation}
\label{eq.1}
s_k\ \rightarrow \hat{s}_k = \frac{s_k}{\sqrt{L}\sigma_k},
\end{equation}
where $\sigma_k$ is the standard deviation of elements in $k$-th subseries.\\
We build $L\times L$ matrices from $\hat{s}_k$ subseries treating them as subsequent matrix rows. The first $L$ subseries build  the first matrix, the  second $L$ subseries build the second matrix, etc. This way, the $(i,j)$ entry of the $n-th$ matrix  $M^{(n)}_{ij}$ ($n=1, ..., [N/L^2]$) reads in terms of time series increments as follows:
\begin{equation}
\label{eq.2}
M^{(n)}_{ij} = \frac{\Delta x_{[(n-1)L+i-1]L+j}}{\sqrt{L}\sigma_{(n-1)L+i}}
\end{equation}

Further steps rely on examination of eigenvalue spectrum properties for the ensemble of matrices built this way and on comparison these properties with some {\it a' priori} known spectrum patterns.
Any distortion of the {\it a' priori} known long range dependence in  the background time series data will be classified as the presence of a new signal immersed in these data. The measurement of this distortion in terms of eigenvalue spectra for matrices built according to the scheme presented above, leads not only to discovery of a new immersed signal but may give us also some information on its parameters.

\section{Data analysis}

Let us look at the beginning, how this  scheme works in the simplest case of white noise chosen as the background, enriched with the sinusoidal signal added at various signal to noise $s/n$ (amplitude) level\footnote{Through this paper $s/n$ has the meaning of ratio between the amplitude $A_s$ of the signal and the standard deviation $\sigma_n$ of the noise data. }. If $s/n =0$, we simply should reconstruct the Wigner semicircle [9].
The crucial observation is here that eigenvalue spectrum $\rho (\lambda)$ for eigenvalues $\lambda$ normalized according to Eq.(1) and obtained from non-correlated data should disappear for $|\lambda|>2$ because of analytic formula describing the Wigner eigenvalue spectrum semicircle:

\begin{equation}
\label{eq.3}
\rho(\lambda)=\frac{1}{2\pi}\sqrt{4-\lambda^2}
\end{equation}

This fact is confirmed numerically in Fig.1 for the ensemble of $10^3$ matrices of size $200\times 200$ constructed from the pure white noise signal. Increasing $s/n$ ratio we see the tails of $\rho(\lambda)$ spectrum becoming much longer and eventually, they exceed the Wigner limit $|\lambda|\leq 2$.\\
We have checked also that obtained results do not qualitatively depend  on period $T$ of the added signal. In plots of Fig.1 period $T \sim 10^6$ was used.

Let us turn to analyze an example of much more complicated but very realistic signal being of wide interest in physics and astrophysics. We examined the background noise data from NAUTILUS experiment [11], i.e. from one of ground placed detectors constructed to detect bursts of gravitational radiation coming from sources like spinning irregular  neutron stars located in our Galaxy or in the Local Group. The idea of NAUTILUS and similar detectors was based on the relative change in macroscopic body sizes (of cylindrical shape for the NAUTILUS experiment) while the hypothetical gravitational wave is passing through it. The relative change in sizes (dimensionless amplitude) up to $10^{-20}$ was expected to be observed by NAUTILUS group. The experiment is no longer running since more modern approaches started data collection like VIRGO, LIGO  or are planned (LISA). Nevertheless, the use of data from NAUTILUS seems to be important to show the efficiency of RM based method  in comparison with the standard Fourier analysis applied to examine such complicated signal. The latter analysis is usually the tool of the first choice in examination of periodic signal in background data what is exactly a case of gravitational wave signal hidden in a background noise of various origin.

We took  $10$ different samples of signals, every
sample containing around $4\times 10^5$ records\footnote{ Samples of respond signal and the simulated noise signal of the detector were received by courtesy of NAUTILUS group}. First, the samples have
been cleared of the obvious technical
distortions (major picks) due to technological reasons, mainly the replacement of liquid argon cooling the cylinder placed inside the apparatus to minimize its thermal noise.
They make time series $n(t)$ of the real signal with the
overall number of data exceeding $L\gtrsim 10^6$. All examined signals were searched  with matrices $200\times 200$, but we checked that results do not significantly depend on the used matrix size.

The eigenvalue spectrum
of matrices built on $n(t)$ time series is plotted in
Fig.2. It makes the characteristic triangle shape very much
different from the semicircle form of Eq.(3). As one could have expected, the time series of shuffled experimental signal $n_{shuff}(t)$ leads to
different eigenvalue spectrum shape - the Wigner one. This is also shown in Fig.2 and confirms only strong correlations in $n(t)$ data. However, there is nothing strange in this observation because detector is adjusted to get maximum sensitivity for specific frequencies of measured signal.

Nevertheless, there is one very striking phenomenon in data analysis shown in details in Fig.3. We compare here eigenvalue spectra for the real detector signal $n(t)$ and the simulated background noise $n_{sim}(t)$ of the detector. The latter signal was used by NAUTILUS group as the reference signal and any deviations from it should be a subject of careful experimental and theoretical analysis. One may see small difference between two spectra, particularly in the tail part. The eigenvalue spectrum for the real signal is wider and heavy tailed comparing with the same spectrum for simulated noise. It is worth to say that this difference between two signals was not noticed in the standard Fourier analysis performed by NAUTILUS group. An importance of such difference, visible only within eigenvalue spectra approach, should be carefully examined, particularly in the spirit of tails modification caused by periodic signal found previously and shown in Fig.1 for simplest signals.

Let us concentrate in the beginning on the simulated detector background noise $n_{sim}(t)$ enriched with a single sinusoidal signal at different $s/n$ ratio. The eigenvalue spectrum for such compound is drawn in Fig.4. We notice that long tails appear in spectrum for such modified signal with local maxima whose position does not depend on the period of added signal. However, the shape of the tail part and its length does depend on the $s/n$ ratio. Simultaneously, the central part of the eigenvalue spectrum (maximum of the spectrum) decreases with decreasing $s/n$ value.
The above observation confirms the role of thick tails in the eigenvalue spectrum and points to the necessity of their exact test for the presence of additional sinusoidal signals, which after all might be hypothetical gravitational waves. Indeed, the gravitational wave can be described as superposition of many sinusoidal signals with varying amplitude and periods [12-15]. It is  described by many parameters defining the mutual angular position and distance of the source with respect to the wave observatory on the ground, orientation of rotation axes, rotation speeds, etc.
Most of these parameters are chosen randomly, since {\it a'priori} we do not know the place in galaxy we can expect the arrival of the signal from. Therefore, the exact shape of expected gravitational wave signal  and the responded signal of detectors is the subject of Monte Carlo simulation [13-15]. Depending on the choice of free parameters, one can get signals varying very much and having different impact on the relative sizes of a probe body.

We took in our further simulations a model of the signal representing
the quadrupole gravitational wave that is emitted by a
freely precessing axisymmetric star [12]. Two extreme detector respond signals with very much different amplitude were generated\footnote{Both signals were offered to us by courtesy of A. Kr\'{o}lak and P. Jaranowski and are based on simulation and analysis described in details in a series of their papers [12-15]} (amplitude ratio of these signals was about one order of magnitude). We will call these respond signals respectively: strong and weak later on. Their eigenvalue spectra are shown in Fig.5 for completeness of our reasoning. They are more complicated than for a single sinusoidal signal but the central pick is well noticed in all cases what reflects the periodic nature of gravitational waves.

Finally, we checked how the presence of such real gravitational wave signal would change the eigenvalue spectrum of the respond signal from the detector if the simulated background noise of detector is also switched on. The outcomes of this simulation is presented in series of Figs.6-8.
The magnitude of gravitational wave signal was kept constant in this simulation, while the level of background noise was the subject of correction $10^k$ times ($k=0,-1,-2,-3)$.  We simulated  this way the eigenvalue spectrum of the entire respond signal expected from the detector in different cases of background noise reduction. This scenario is expected to come true in ground based experiments in near future.

It is seen that in approach based on eigenvalue spectrum, the strong signal of gravitational wave passing through the Earth is going to be detected already at the present sensitivity of NAUTILUS detector $ (k={0})$  (see Figs.6-8). The weaker respond signal can be found if the background noise level is reduced $10^{-2}$ times (see Fig.8). The positive respond signal of gravitational wave will most easier be seen in both cases by modification of the tail and head part in the eigenvalue spectrum as indicated in Figs.7-8.

\section{Conclusions}
Concluding, we presented a novel approach in detection of repeatable signals in noisy data. Our analysis was focused on the real noisy data obtained from NAUTILUS gravitational wave experiment, as an example. In order to get in future unambiguous experimental evidence of
gravitational waves, one has to apply independent methods, based on
various theoretical philosophies. New approach we proposed is
based on RM techniques. It revealed an existing subtle difference
between the simulated detector noise and truly recorded background
signal. This difference was not evident so far with the use of other
techniques like Fourier analysis and  has to be eliminated if very weak gravitational waves
signal is going to be
crosschecked by independent and diversified methods of data
analysis. Moreover, we provided also the eigenvalue spectrum of the entire respond signal expected from ground based detector in future experimental activities, by diminishing its background noise level. We believe that the good performance of our method in the latter case is a good predictor of further applications in many other areas as well.

\section*{Acknowledgements}

Authors would like to thank  A. Kr\'{o}lak for helpful discussion
and NAUTILUS group for delivery the samples of experimental data and the background signal of the NAUTILUS detector for the purpose of this elaboration. Authors thank also M. Daszkiewicz for suggesting the problem and contacting us with gravitational wave researchers.

\newpage

\begin{figure}
\begin{center}
 \includegraphics[scale=1.3]{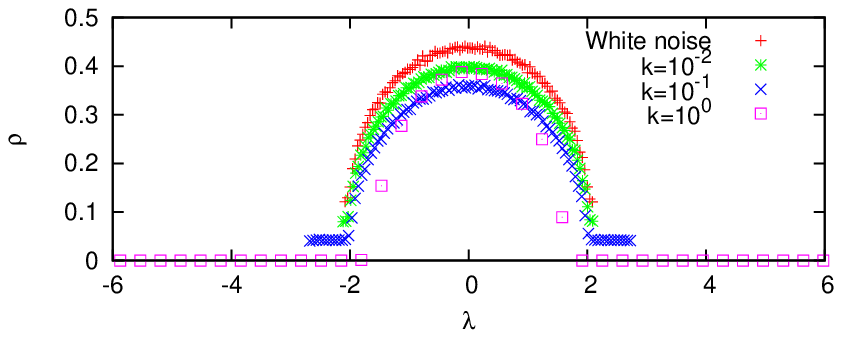}
\end{center}

 \caption{Example of eigenvalue spectra for white noise enriched with sinusoidal signal at various signal to noise ratios ($s/n=10^k, k=0,-1,-2)$. Simulation done from $10^3$ matrices $200\times  200$.}
 \label{fig.1}
\end{figure}

\begin{figure}
\begin{center}
 \includegraphics[scale=1.3]{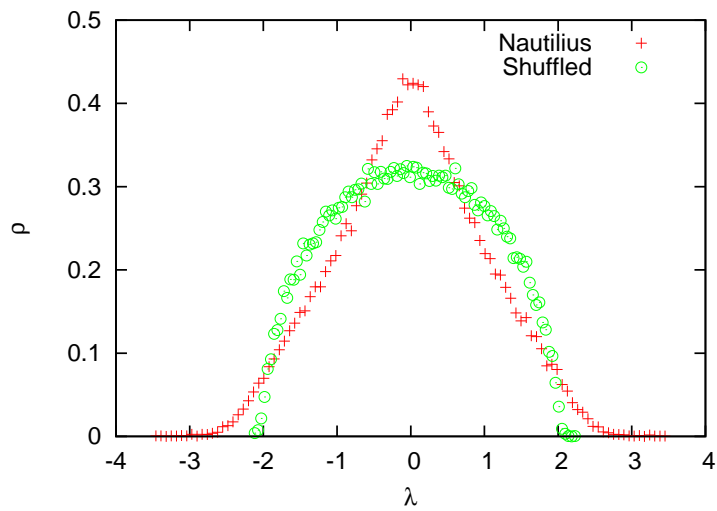}
\end{center}
 \caption{Eigenvalue spectrum for original data $n(t)$ from NAUTILUS and its shuffle $n_{shuff}(t)$}
 \label{fig.2}
\end{figure}

\begin{figure}
\begin{center}
 \includegraphics[scale=1.3]{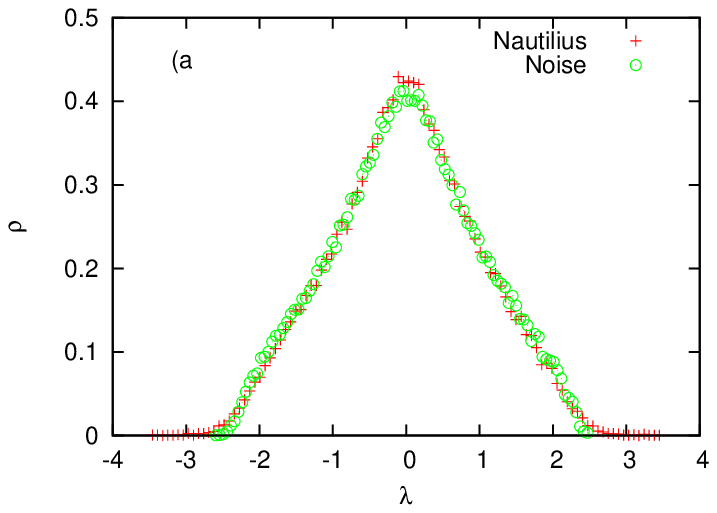}
 \includegraphics[scale=1.3]{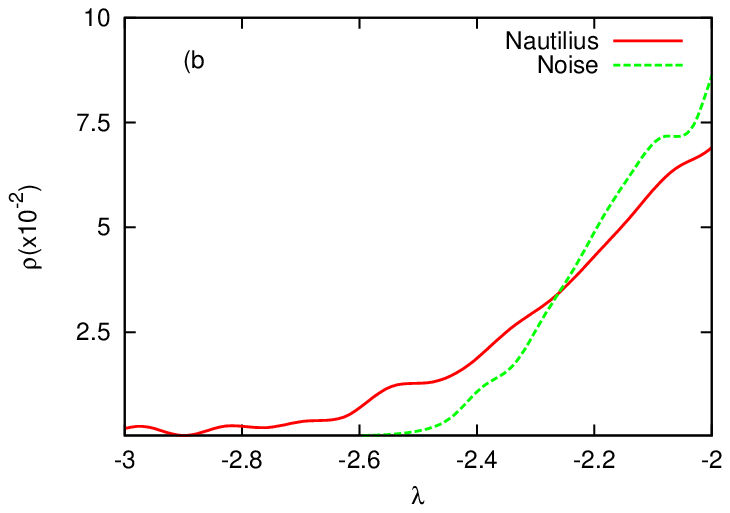}
\end{center}
 \caption{Comparison between two eigenvalue spectra calculated for the original detector's signal and the simulated background noise of detector (a. Difference in the tail part of two spectra is noticed (b.}
 \label{fig.3}
\end{figure}

\begin{figure}
\begin{center}
 \includegraphics[scale=0.85]{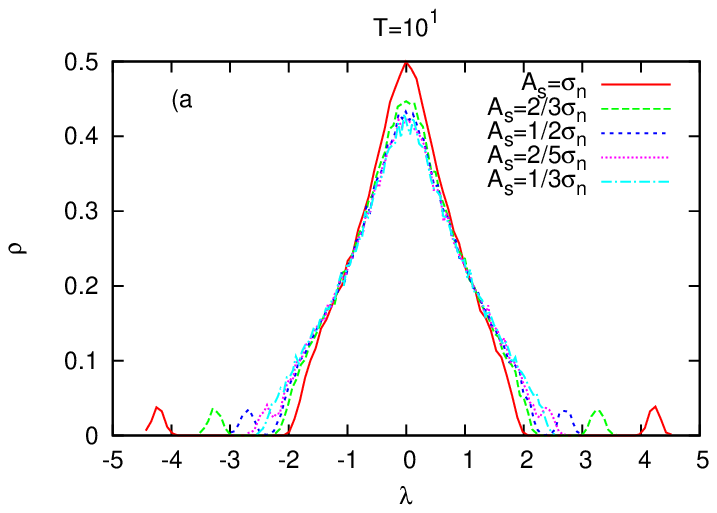}
\includegraphics[scale=0.85]{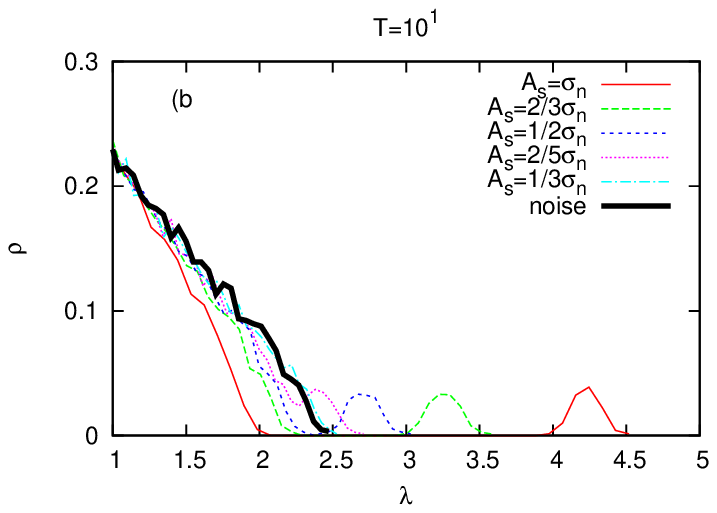}
\includegraphics[scale=0.85]{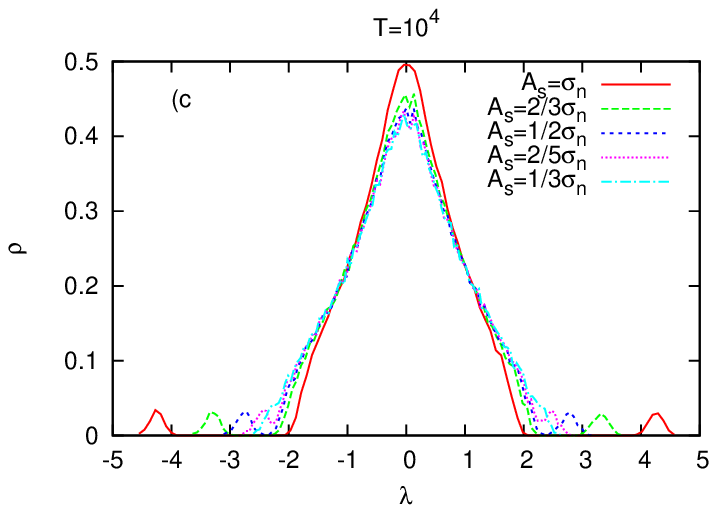}
\includegraphics[scale=0.85]{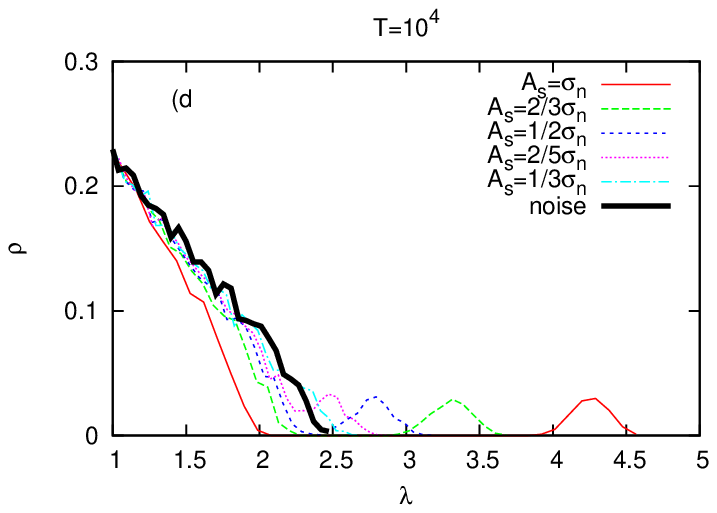}
\end{center}
 \caption{Simulated expected eigenvalue spectra for the NAUTILUS background noise with added sinusoidal signal at various $s/n$ ratios for two different periods: $T=10$ (a) and $T=10^4$ (c). Figs.4b, 4d show the zoomed part of both cases, indicating for comparison the actual detector noise level.}
 \label{fig.4}
\end{figure}

\begin{figure}
\begin{center}
 \includegraphics{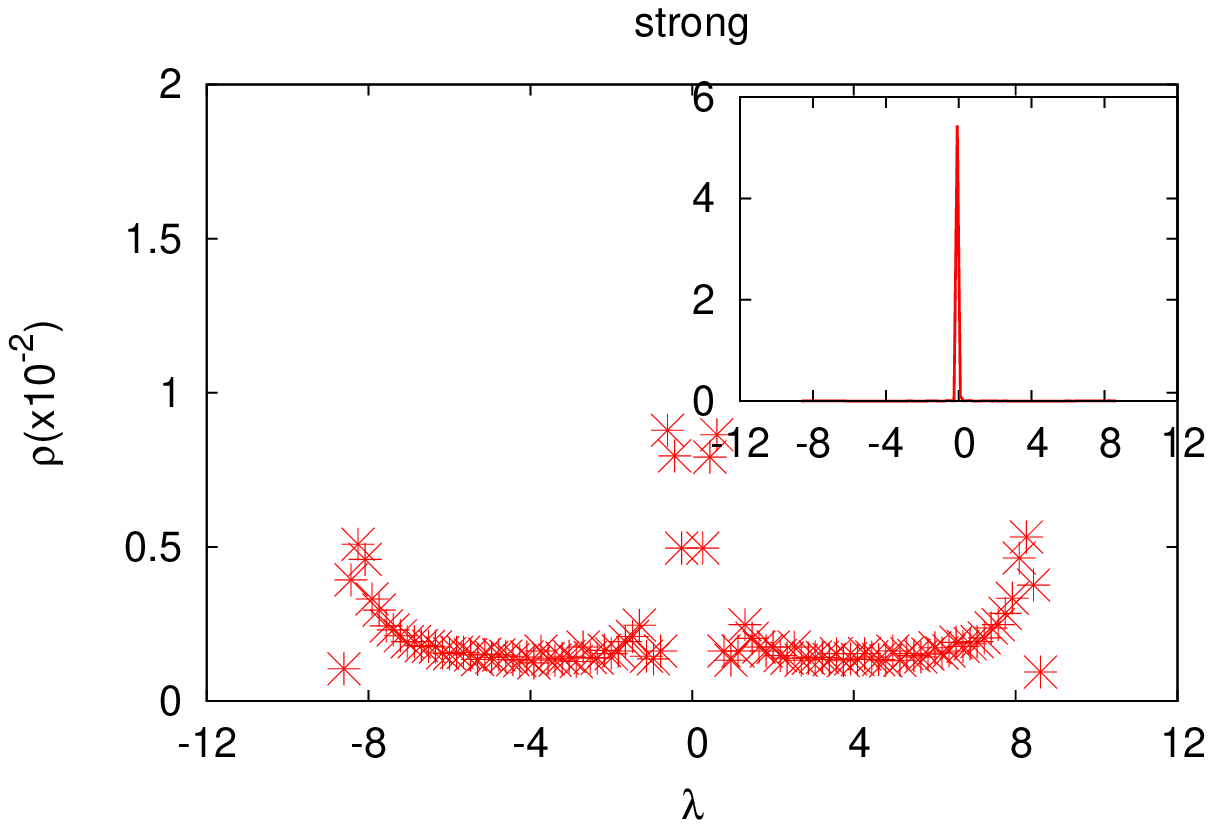}
\includegraphics{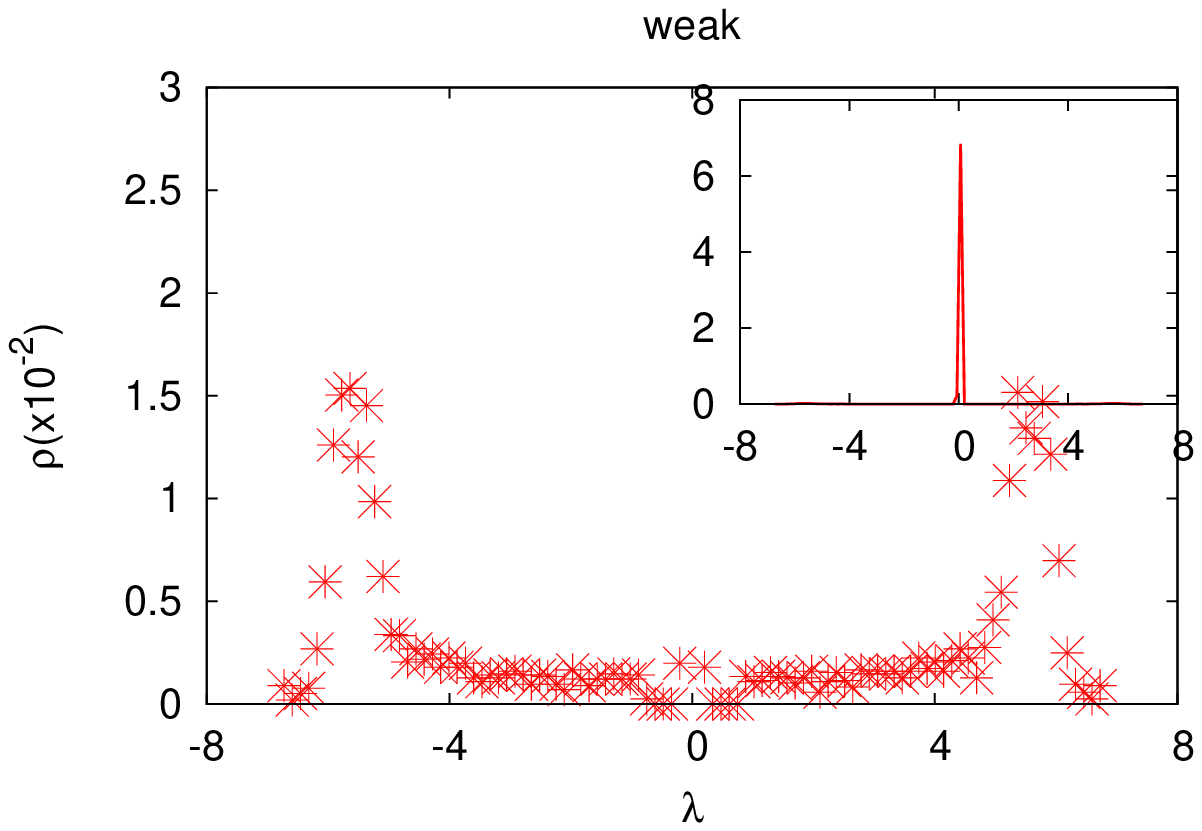}
\end{center}

 \caption{Eigenvalue spectra of the detector's respond signal for the case of strong  and weak  gravitational wave and neglected background noise. Shape of the tail part is magnified.}
 \label{fig.5}
\end{figure}

\begin{figure}

\begin{center}
 \includegraphics[scale=1.3]{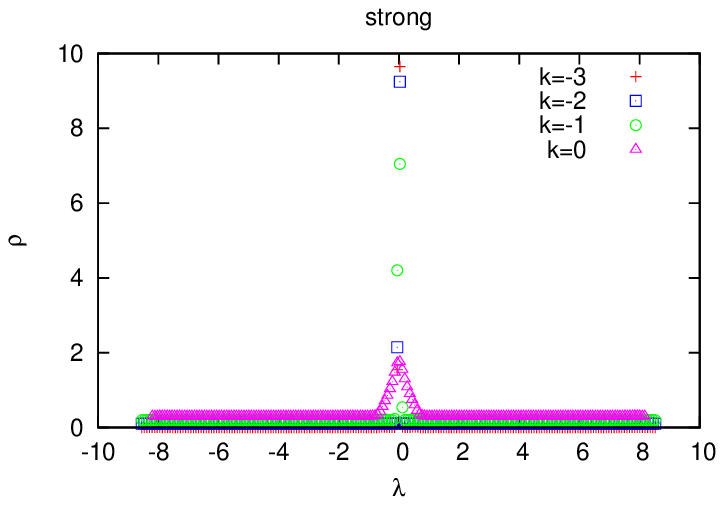}
\includegraphics[scale=1.3]{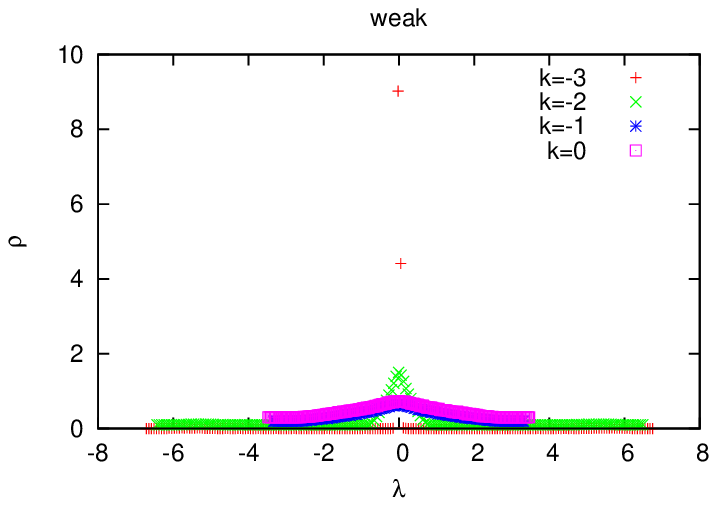}
\end{center}
 \caption{Eigenvalue spectra of the detector's respond signal for the case of strong  and weak  gravitational wave with the simulated background noise added. The magnitude of the tested signal is kept constant, while the background noise is being changed $10^k$ times, ($k=0,-1,-2,-3$) accordingly.}
 \label{fig.6}
\end{figure}

\begin{figure}
 \begin{center}
 \includegraphics[scale=1.3]{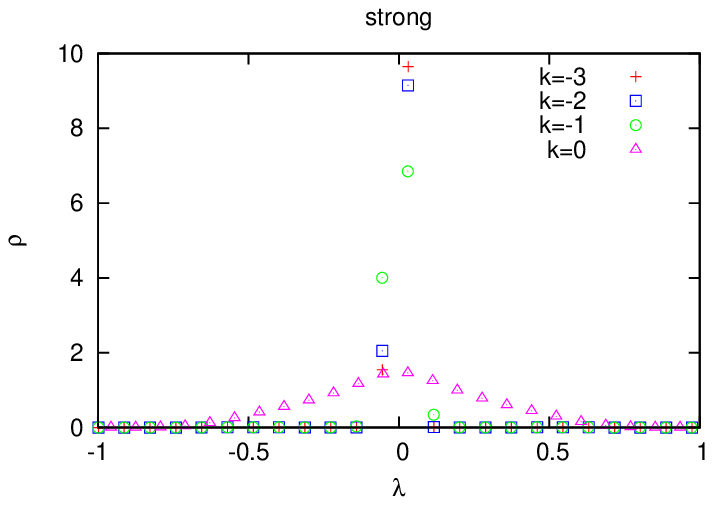}
 \includegraphics[scale=1.3]{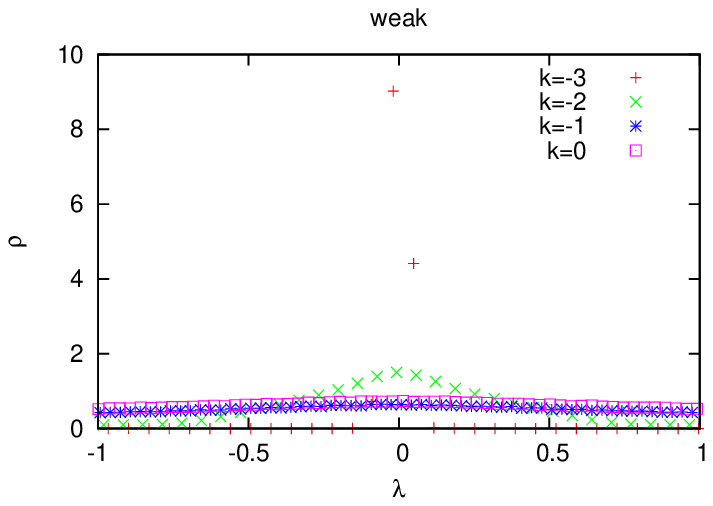}
\end{center}
 \caption{Magnified central part of the spectra from Fig.6.}
 \label{fig.7}
\end{figure}

\begin{figure}
 \begin{center}
 \includegraphics[scale=1.3]{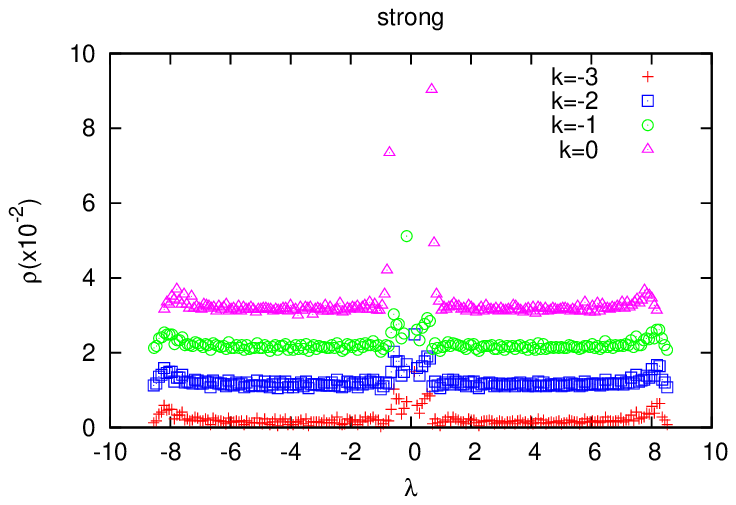}
\includegraphics[scale=1.3]{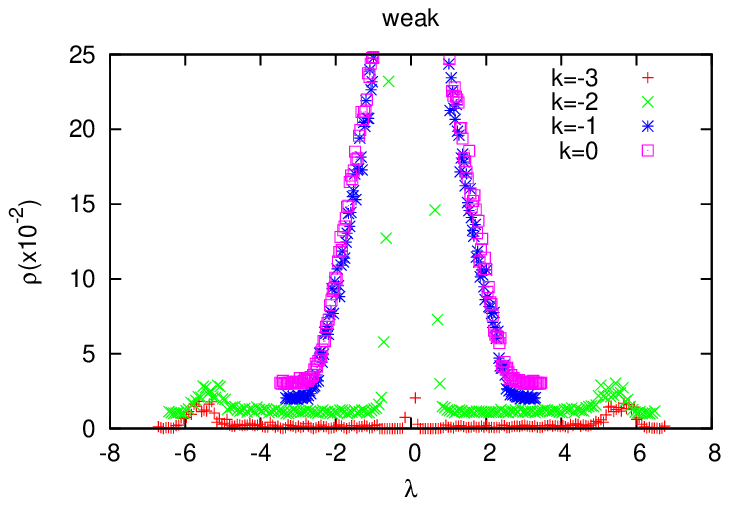}
\end{center}
 \caption{Magnified and artificially shifted vertically the tail part of the spectra from Fig.6.}
 \label{fig.8}
\end{figure}

\end{document}